\documentclass[aps,showpacs,pra,twocolumn,groupedaddress]{revtex4-1}

\usepackage[dvips]{graphicx}
\usepackage{epsfig}
\usepackage{subfig}
\usepackage{dcolumn}
\usepackage{bm}
\usepackage{amsmath}
\usepackage{pdfsync}
\usepackage{amssymb}
\begin{document}

\preprint{PBK-Physics}

\title{Loading characteristics of a microscopic optical dipole trap}

\author{P. Kulatunga}
\email[]{kulatunga@hws.edu}
\altaffiliation{Hobart \& William Smith Colleges}
\affiliation{Hobart \& William Smith Colleges, Geneva, NY 14456, USA}
\author{T. Blum}
\affiliation{Hobart \& William Smith Colleges, Geneva, NY 14456, USA}
\author{D. Olek}
\affiliation{Hobart \& William Smith Colleges, Geneva, NY 14456, USA}
\date{June 01, 2010}

\begin{abstract}
We report on an investigation of loading characteristics of deep microscopic dipole traps. The dipole trap is loaded from a low density magneto optical trap (MOT) containing $\approx 5\times 10^{6}$ atoms.  We determine the loading parameters that  maximize the trapped atom number for a trap of waist 5 $\mu$m with trap depths ranging from 3.5 mK to 10 mK. We determine the optimal trap loading conditions and the loading rates, loss coefficients and temperature of the trapped atoms under these conditions.  We show that it is possible to load a few hundred to thousand atoms in dipole traps of depth 3.5 mK to 8.5 mK under the optimal loading conditions.
\end{abstract}


\maketitle

\section{Introduction}
\indent In the last two decades trapped ultracold neutral atoms have been studied extensively and continue to be of interest to many investigators in atomic, molecular and optical physics.  Of the many trapping techniques available, the far-off resonance optical traps (FORT), requires no magnetic field while providing a tight, nearly conservative potential characterized by low optical excitation rates \cite{7Miller:PhysRevA93}. Ease of control of trapping parameters and long lifetime in these traps has allowed for applications in precision measurements, to realize all optical Bose-Einstein Condensates (BEC), and use in parity non-conservation and $\beta$ decay asymmetry measurements \cite{4Romalis:PhysRevA2001,  Chapman:PhysRevLett2001, Kinoshita:PhysRevA:2005, Gwinner:HypInt:2007, Crane:PhysRevLett2001}.  Dipole traps are also used for studies in photo-association, such as in investigations of Feshbach resonances in ultracold gases and Efimov physics of few body quantum systems \cite{Lett:Julienne:RevModPhys2006, Kramer|Nature2006}. Recent work involving neutral atom quantum computing and cavity quantum dynamics have required trapping few to a single atom in a microscopic dipole trap\cite{Saffman:PhysRevA:2008, 18Yavuz:PhysRevLett:06, Puppe:PhysRevLett:2007}. Also of interest are traps that maximize the density of atoms, such as those required for investigations in quantum multiple scattering of photons and in studies that have been  proposed to observe strong localization of light \cite{Kaiser:JModernOptics:2009}. 

The optical dipole force is a  conservative force that does not contribute to cooling. Therefore these  traps are always loaded from a pre-cooled sample of atoms. Most often the traps are  designed to optimize loading from a sample of magneto-optically trapped (MOT) atoms to maximize the trapped number. Dynamics of loading dipole traps characterized by waist of 10 $\mu$m and larger from magneto-optical trap (MOTs) have been investigated \cite{7Miller:PhysRevA93, 20Kuppens:PhysRevA2000, DCho:PhysRevLett:1999, Ohara:PhysRevA:2001}.  A comparison of loading dynamics of CW and pulsed dipole traps of trap waist 16 $\mu$m has also been reported \cite{Sukenik:PhysRevA2008}. It has been shown that the trapping efficiency is dependent on the trap volume as well as the density and the temperature of the MOT atoms \cite{20Kuppens:PhysRevA2000}. Loading of $\approx$ 1 mK deep microscopic dipole trap, having  waist 0.7 $\mu$m to 11 $\mu$m had been  investigated over many orders of loading rates  in the regime of trapping one to few atoms \cite{Schlosser:PhysRevLett02}. These studies have revealed interesting dynamics and physics that govern the loading of these shallower microscopic dipole traps \cite{11Schlosser:Nature01}.

Our particular motivation is to study coherent multiple scattering of photons to investigate strong localization in a high density sample of trapped ultracold atoms. The atomic densities desired for these investigations are approximately three orders of magnitude higher than that is achieved in a MOT \cite{Kaiser:JModernOptics:2009}.  High density, far-detuned dipole traps (having a large waist) are now investigated as possible means of reaching this limit \cite{Kaiser:JModernOptics:2009, QESTHavey}. However, a sample of atoms trapped and cooled to the ground state of a microscopic dipole trap is also a promising approach to consider under conditions which optimizes the number of trapped atoms.

In this report we present the results of loading dynamics of deep, microscopic traps from a low density MOT containing $\approx 5 \times 10^{6}$ atoms. Here we define a large trap as one characterized by a transverse waist $w_{0}$ such that the Rayleigh range of the focus spot $z_{R}=\pi w_{0}^{2}/\lambda \geq r_{MOT}$, where $r_{MOT}$ is the MOT radius. During the loading process the MOT and the dipole trap is over-lapped, and for large traps the dominant axial confining potential is provided by the MOT \cite{Ohara:PhysRevA:2001}. In the limit that $z_{R}\ll r_{MOT}$, as is the case for a microscopic dipole trap the dominant axial and transverse confining potential is the FORT potential. Therefore it is possible, the loading dynamics of the larger traps differ from the latter.  Transfer of atoms into the dipole trap is a dynamical process that is dependent on many competing processes such as initial MOT density, trap losses due to density dependent collisions, trap depth, trap alignment with respect to the MOT, etc. The lifetime of the atoms remaining in the dipole trap subsequent to the loading process is determined by the background collision rate, scattering rate of trap laser photons as well as  alignment stability and intensity noise of the trapping laser.

It has been shown that to achieve efficient loading of a dipole trap from a MOT the loading parameters, the hyperfine repumping intensity and the MOT detuning must be changed during the loading stage \cite{20Kuppens:PhysRevA2000, Sukenik:PhysRevA2008}. However, to our knowledge there has not been a comprehensive study of loading dynamics, temperature, lifetime and loss rates of atoms in deep microscopic dipole traps. As the trap size is reduced the loading efficiency is also reduced, but little is known of, and how loading parameters can be optimized for efficient loading of a microscopic dipole trap. We report on the loading dynamics of a dipole trap of waist of $\approx 5 \mu$m of depth ranging from 3.5 mK to 10 mK under various loading conditions. We measure the optimal loading conditions and obtain the loading rates and loss coefficients under these conditions. 
\section{Trap Potential\label{potential}}
The trapping  potential is defined by the Gaussian profile of the trapping beam at the focus. The trapping potential is given in terms of the radial and the longitudinal coordinates $(\rho,z)$ as,
\begin{equation}
U(\rho,z)=U_{0}\frac{exp\left [-2\rho^{2}/w(z)^{2}\right ]}{1+\left(z/z_{R}\right )^{2}}.
\end{equation}
Here $U_{0}$ is the maximum trap depth that depends on trapping laser power, the beam waist, and the detuning from the atomic resonance. And the beam waist as a function of the longitudinal coordinate $z$ is given by $w(z)^{2}=w_{0}(1+z/z_{R})^{2}$, where $z_{R} =w_{0}^{2}\pi/\lambda$ is the Rayleigh length of the focus spot \cite{20Kuppens:PhysRevA2000}. For a linearly polarized FORT, the maximum trap depth for an alkali-metal atom is given by 
\begin {equation}
U_{0}=\frac{\hbar \gamma I_{0}}{24 I_{s}}\left[ \left(\frac{1}{\Delta_{1/2}}+\frac{2}{\Delta_{3/2}}\right)\right],
\end{equation}
where $\gamma = 2\pi\times 5.98$ MHz is the natural line-width of Rb, $I_{s}$ is the saturation intensity and $I_{0}$ is the peak intensity. The detunings $\Delta_{1/2}$ and $\Delta_{3/2}$ represent the difference between the laser frequency and the $D_{1}$ and $D_{2}$ transition frequencies expressed in multiples of $\gamma$ \cite{DCho:PhysRevLett:1999, 20Kuppens:PhysRevA2000}.

\section{Experiment}
\begin{figure}[h]
\includegraphics[width = 3.5in]{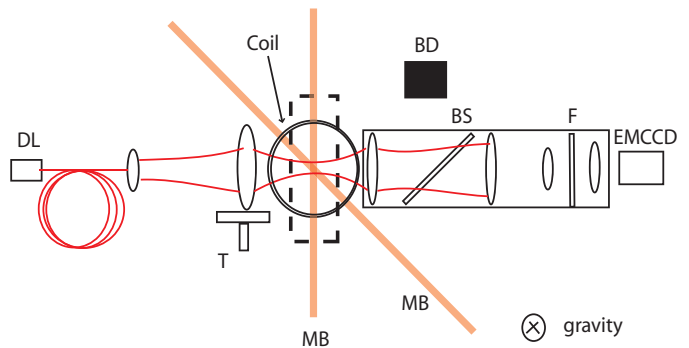}
\caption{(color online) Experimental setup (viewed from top). DL - Dipole laser, MB - MOT beams, BS - Dichroic beam splitter, F - 780 nm bandpass filter, T - $xyz$ translator, BD - beam dump.} {\label{setup}}
\end{figure}

The experimental setup shown in Fig.\ \ref{setup} consists of a 6 beam MOT loaded from the background vapor, optics for tight focusing of the FORT laser, fast detection system and computer interface for control and data acquisition. The unconventional  x and y beam configuration of the MOT setup shown in Fig.\ \ref{setup} maximizes the optical access to tightly focus the FORT beam and the collection solid angle for  fluorescence detection using optics placed outside the vacuum chamber. Two external cavity diode lasers stabilized to $^{85}$Rb atomic lines at 780 nm are used for the MOT. Both MOT lasers are independently controlled by two acousto-optical modulators (AOM). The main MOT trapping beam (and the hyperfine) is single passed through the AOM. In this configuration the MOT detuning was restricted to a very narrow frequency range of 6 MHz. The first order beam is coupled to a polarization maintaining single mode fiber that is used for spatial filtering as well as for ease of delivery. The first diffracted order from the hyperfine AOM is expanded and aligned directly to the trap along the two horizontal trapping beams. During the FORT loading stage the main trapping/cooling beam can be detuned by 6 MHz at the expense of decreasing the power at the trap due to reduced coupling to the fiber resulting from the transverse motion of the 1st order beam. The maximum magnetic field gradient of the quadrupole field along the strong axis is 7 G/cm. Under normal MOT loading conditions the field gradient was set to 4 G/cm.
The MOT laser is detuned by 12 MHz to the red of the  $5^{2}S_{1/2} F=3 \rightarrow 5^{2}P_{3/2}F^{\prime} =4$ transition. The trapping power is divided between three retroreflected beams, each of diameter 8.0 mm resulting in trapping and cooling intensity of $I_{MOT}= 28$ mW/cm$^{2}$. The hyperfine repump laser is tuned to $5^{2}S_{1/2} F=2 \rightarrow 5^{2}P_{3/2}F^{\prime} =3$ transition with a nominal detuning of $\Delta_{R}=0$ unless otherwise stated. The repumping laser is divided and co-propagated along the horizontal MOT beams.  The repumping intensity maximum is $I_{R}= 6$ mW/cm$^{2}$.

The FORT laser is an IPG Photonics fiber laser operating at a wavelength of 1064 nm.  It delivers  a maximum of 10.2 W  through a single mode polarization maintaining fiber. Short term intensity variations are less than 0.5\%rms (1 kHz - 100 MHz) and long term (over 5 hours) variations are less than 2\%rms. The fiber output is a well collimated Gaussian beam of M$^{2} <1.3$. The fiber output is expanded, re-collimated and focused into the chamber with focal length 100 mm achromatic doublet objective lens. This objective lens is mounted on a precision xyz translator. All the lenses in the optical system are used on axis, and well within the manufacture suggested f/\# for diffraction limited performance. The calculated diffraction limited focus radius is 5 $\mu$m. However in our calculations of the trap depth, a spot size of 5.6 $\mu$m is used to account for aberrations due to windows and alignment imperfections.  The actual trap size was verified by parametrically oscillating the FORT intensity and  measuring the atom loss as a function of the modulation frequency.  These measurements were inherently noisy due to low atoms numbers captured in the trap. Additionally only the losses at twice the longitudinal frequency was observed due to limited maximum attainable modulation frequency of the laser intensity. With this measurement we estimate the upper bound of the trap waist at 8 $\mu$m.

\subsection{Trap loading and detection}
The MOT is loaded for 3 s with maximum trapping intensity $I_{T}= 28$ mW/cm$^{2}$, repump intensity $I_{R}=4.2$ mW/cm$^{2}$, and a field gradient of 4 G/cm. With these MOT parameters approximately $5\times10^{6}$ atoms are loaded giving a MOT density of $\approx 2 \times 10 ^{10}$ atoms/cm$^{3}$.  We attribute the low atom number in the MOT to the unconventional trap configuration, and the small beam size of the the MOT lasers used to minimize light scattered into detection optics. During the FORT loading sequence the hyperfine intensity is reduced and the MOT is detuned for a variable period of time, which we call the loading time. At the end of the loading period the repumping light, the trap lasers and the magnetic field are turned off. Atoms trapped in the FORT are held for a period of at least 75 ms to allow the MOT atoms to dissipate from the detection region before the FORT atoms are detected. The number of atoms is determined by measuring the fluorescence of the atoms in the optical molasses of the MOT beams \cite{Townsend:PhysRevA1995}. This is achieved by extinguishing the FORT laser and simultaneously turning the MOT lasers on. The trapping laser is turned on at full power at a detuning of $-2\gamma$ and the repump laser at 2 mW for 0.5 ms, during which the fluorescence is detected by an electron multiplying CCD (Andor iXon EMCCD) camera. The collection solid angle of the detection system is 0.2 steradians, giving a fluorescence collection efficiency of 1.6\%.  As shown in Fig.\ \ref{setup} we observe in the longitudinal direction, along $-z$, hence only the transverse shape of the cloud of trapped atoms is observed.

The number of atoms captured in the FORT is measured as function of the loading time. During the loading time the MOT is held at a fixed detuning of $-2.5\gamma$  and the hyperfine repump laser is attenuated over a range of values to maximize FORT loading. The loading curves and lifetime curves are measured for trap depths ranging from 3.5 mK to 8.7 mK at a fixed, calculated trap waist of 5 $\mu$m. With its Rayleigh length, $z_{R}=\pi w_{0}^{2}/\lambda \ll r_{MOT}$, the trap can be considered microscopic as defined earlier. We have investigated the loading curves at different repump intensities to determine the optimum loading time at which the number of atoms trapped in the FORT is maximized for a range of trap depths. The intensity of hyperfine repump laser $(I_{R})$, trap intensity $(I_{M})$, detuning of trap laser and repump laser, position of the FORT  relative the MOT center, density of the MOT in combination form a large parameter space to investigate optimal loading conditions. We have measured the loading characteristics under several parameters, as function of trap depth, hyperfine attenuation,  position of the FORT relative to the MOT and the polarization of the FORT laser beam. We also measure the atom temperature and lifetime in the FORT as a function of the trap depth. 
\subsection{Dynamics of microscopic trap loading}
\subsubsection{Loading}
We investigated the loading as a function of trap depth, hyperfine power, loading time and the MOT density to find the optimal number of atoms that can be transferred to the microscopic dipole trap from the FORT. The FORT depth was varied from 3.5 mK to 8.6 mK while keeping the trap waist fixed. The duration of the time that the FORT is overlapped with the MOT with the primary MOT light, the hyperfine repump light and the magnetic field gradient is defined as the loading time. The following rate equation has been shown to model loading of a FORT from a MOT \cite{20Kuppens:PhysRevA2000, Schlosser:PhysRevLett02},
\begin{equation}
\label{eq:loading}
\frac{dN}{dt} = R_{0}exp(-\gamma_{MOT}~t)-\Gamma_{L}N-\beta_{L}^{\prime}N(N-1).
\end{equation}
Here $R_{0}$ is the initial loading rate and $\gamma_{MOT}$ characterizes the loss of atoms from the MOT due to the changes to the MOT loading conditions during the loading stage. The parameters $\Gamma_{L}$ and $\beta^{\prime}_{L}$ are the loss rates due to collisions with the background (proportional to $N$) and loss rate due to two body collisions (proportional $N(N-1)$) respectively. Subscript $L$ is used in Eqn.\ \ref{eq:loading} to differentiate the loss rates during loading when the atoms are under the influence of the MOT light from the loss rates during storage in the FORT, in the absence of the MOT light and the field gradient. The prime on $\beta^{\prime}$ refers to the atom number loss rate and differentiates from the density loss, $\beta$. 

\begin{figure}[h]
\centering
\includegraphics[width=60mm, height=40mm]{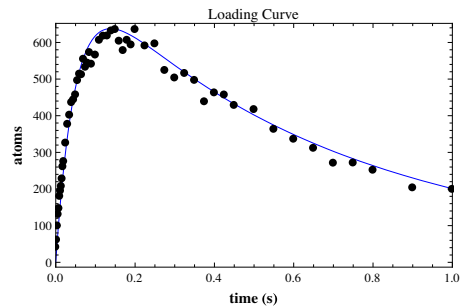}
\caption{Number of atoms in the trap during loading, FORT depth of 4 mK, waist of $\approx 5~\mu$m. Main MOT intensity $I_{M}=28$ mW/cm$^{2}$, hyperfine repump intensity $I_{R}=50$ $\mu $W/cm$^{2}$. } {\label{fig:loadcurve1}}
\end{figure}

We can expect the number of atoms in the FORT to increase linearly with time at small loading times as $N(t) = R_{0}t$. At larger loading times the number will roll-off as the MOT loses atoms due to reduced repump intensity. As the number of atoms in the FORT increases, losses will begin to define the tail of the loading curve. The dynamics of the microscopic trap loading is investigated by measuring the number of atoms $N(t)$ in the FORT as a function of the loading parameters. Shown in Fig.\ \ref{fig:loadcurve1} is the measured number of atoms in the FORT as a function of the loading time and the fit (solid line) to Eqn.\ \ref{eq:loading} for trap depth of 4 mK for $I_{R}=50~\mu$W/cm$^{2}$. Here each data point is an average of 10 acquisitions, where each acquisition is one loading cycle composed of MOT loading, FORT loading, holding and detection. 
All four parameters that define the loading can be extracted by fitting the data to the numerical solution of Eqn.\ \ref{eq:loading}. The number of atoms in the FORT $N(t)$ is measured in loading time increments of 2 ms from the beginning of the loading stage up to 20 ms, in 5 ms increments from 20 ms to 100 ms, and subsequently in larger increments up to loading times exceeding 1 s. As we expect the number of atoms in the FORT to increase linearly with time for small $t$, we use the initial 20 ms to estimate the loading rate $R_{0}$ from the slope of the loading curve at small loading times. And we fix $R_{0}$ and fit the numerical solution of Eqn.\ \ref{eq:loading} to extract the loss coefficients $\Gamma_{L}$ and $\beta^{\prime}_{L}$. While this provided  a check for consistency, the values obtained for the parameters were not very much different when $R_{0}$ was not constrained. For the loading curve shown in Fig.\ \ref{fig:loadcurve1} values obtained are  $\Gamma_{L} = 1.25$ s$^{-1}$ and $\beta^{\prime}_{L} = 4\pm1 \times 10^{-4}$ (atoms s)$^{-1}$.

The first term of Eqn.\ \ref{eq:loading}, the initial loading rate $R(\tau) = R_{0}exp(-\gamma_{MOT}~\tau)$ can be measured directly by allowing the MOT to dissipate for a variable length of time $\tau$ before the beginning of the loading time \cite{20Kuppens:PhysRevA2000}. We allowed the MOT to dissipate and measured $R(\tau) = R_{0}exp(-\gamma_{MOT}~\tau)$ at several repump intensities. However the values for $R_{0}$ and $\gamma_{MOT}$ from these measurements underestimated the values obtained from the loading curves by as much as 50\%.  We assume that this is due to a systematic error in the method we used to dissipate the MOT, therefore we did not use the values obtained from this method in evaluating the loss coefficients or the loading rates that we report below.
\subsubsection{\label{holding}Holding}
Subsequent to the loading stage, the atoms are held in the FORT for a period of time we call the holding time. During the holding time atoms are lost from the FORT due to collisions with background atoms and through two body collisions. These loss rates define the lifetime of the FORT according to the following rate equation \cite{20Kuppens:PhysRevA2000},
\begin{equation}
\label{eq:holding}
\frac{dN}{dt}  = -\Gamma_{H} N-\beta^{\prime}_{H}N(N-1).
\end{equation}
Here $\Gamma_{H}$ and $\beta^{\prime}_{H}$ are the loss rates in the absence of MOT light. The subscript $H$ denotes loss rates during holding  in the absence of any MOT light or the magnetic field gradient. For each trap depth investigated, holding curves or lifetime measurements were obtained. Subsequent to loading the FORT for an optimal loading time, atoms were held in the FORT for a minimum of 75 ms for the free MOT atoms to disperse before the FORT is extinguished and the atoms are detected with the MOT light. The number of atoms remaining in the trap as a function of the FORT holding time is fitted to the analytical solution of Eqn.\ \ref{eq:holding} to extract the loss coefficients $\Gamma_{H}$ and $\beta^{\prime}_{H}$.
 
We also measured the FORT lifetime in the presence of the primary MOT light  but in the absence of the field gradient. Such a measurement of the holding curve allows us to determine the loss rates, independent of the loading rate \cite{20Kuppens:PhysRevA2000}. In the absence of the magnetic field the FORT loading rate must decrease. It is not identically zero as it is possible to load from the optical molasses. However the influence of the MOT light on the FORT atoms remain, and allows to investigate the effect of the MOT light on the atom loss. This loss curve must still be defined by  Eqn.\ \ref{eq:holding}, however with coefficients $\Gamma_{H}$ and $\beta^{\prime}_{H}$ replaced by $\Gamma_{H^{\prime}}$ and $\beta^{\prime}_{H^{\prime}}$. The loss coefficients are extracted from fitting the data to the analytical solution of Eqn.\ \ref{eq:holding} with coefficients $\Gamma_{H^{\prime}}$ and $\beta^{\prime}_{H^{\prime}}$. The number of atoms in the FORT subsequent to the loading period is measured as a function of time that the atoms are held in the FORT while exposed to the trapping and the repump light without the magnetic field. In essence the FORT lifetime in the presence of the optical molasses.

\begin{figure}[h]
\centering
\includegraphics[width=60mm, height=40mm]{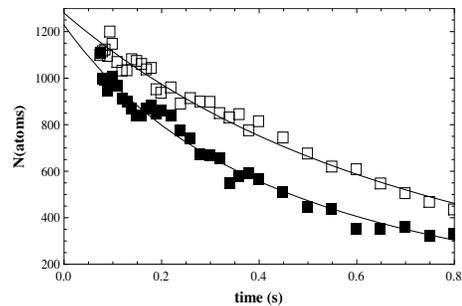}
\caption{Number of atoms as a function of holding time, $\square$ without primary MOT and repump light, and $\blacksquare$ with primary MOT and repump lasers on at $I_{MOT}=28$ mW/cm$^{2}$ and $I_{R}=100$ $\mu$W/cm$^{2}$during the holding time. FORT depth of 8.5 mK, waist of 5.6 $\mu$m.} {\label{fig:lifetimeAB}}
\end{figure}

Shown in Fig.\ \ref{fig:lifetimeAB} is the holding curves with and without primary MOT and repump light for a trap depth of 8.5 mK. The $(\square)$ represents the data in the absence of the molasses and $(\blacksquare)$ the number of atoms in the presence of the MOT light. The solid line is a fit to Eqn.\ \ref{eq:holding}. Without the MOT light present  $\beta_{H}^{\prime} = 3 \times 10^{-4}$ (atoms s)$^{-1}$, and with the MOT light during holding $\beta^{\prime}_{H^{\prime}} = 11 \times 10^{-4}$ (atoms s)$^{-1}$. These values are obtained with $\Gamma_{H}$ and $\Gamma_{H^{\prime}}$ fixed at $1.05$ (atoms s)$^{-1}$.  The values $\Gamma_{H,H^{\prime}}=1.05$ (atoms s)$^{-1}$ is consistent with the lifetime of the FORT due to background pressure in the chamber \cite{Sukenik:PhysRevA2008}. 

The initial loading rates, the loss coefficients during loading, and the loss coefficients during holding were measured under varying loading conditions and for different trap depths. The initial loading rate $R_{0}$ was obtained from the slope of the loading curve at small loading times. Uncertainty of this value was also estimated from the linear least squares fit.  The numerical solution to Eqn.\ \ref{eq:loading} was fitted to the data with $R_{0}$ constrained to the value obtained from the slope of the loading curve. Additionally, the loss coefficient $\Gamma_{L}$ and $\beta_{L}^{\prime}$ were also evaluated from the measurements of the holding curve in the presence of the MOT light field. The values of $\gamma_{MOT}$ extracted from the loading curves obtained at different repump intensities $I_{R}$ were observed to vary by  30\%, with higher values, as expected, obtained for lower repump values. The fits to the loading curves obtained at different trap depths but at a fixed repump power gave $\gamma_{MOT}$ values that changed by 18\%. The error bars on the results presented below represent the statistical variations between different data sets taken under fixed conditions, and the spread in the values obtained when the fixed parameters $R_{0}$ and $\gamma_{MOT}$ are varied within the limits stated. In the following section we discuss the results obtained under different loading conditions.


\subsection{Loading as a function of Hyperfine intensity}
\begin{figure}[h]
\includegraphics[scale=0.5]{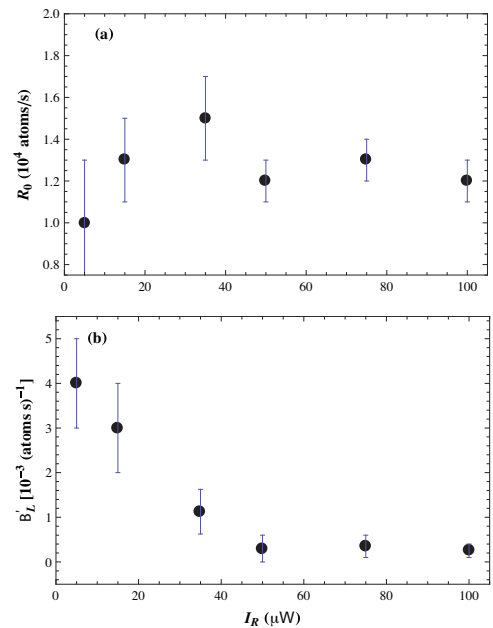}
\caption{(a) Loading rate $R_{0}$ and (b) loss coefficient $\beta_{L}^{\prime}$ as a function of repump power $I_{R}$. MOT intensity $I_{T} = 28$ mW/cm$^{2}$, field gradient 4 G/cm. FORT depth at 4 mK, P = 2.57 W, $w_{0}= 5.59~\mu$m. Error bars are statistical and are estimated from the fits. } {\label{fig:r0andbeta}}
\end{figure}

A common technique used to improve loading into a FORT is to reduce the hyperfine repump intensity.  In Fig.\ \ref{fig:r0andbeta} we show the dependence of $R_{0}$ and $\beta_{L}^{\prime}$ on the repump intensity during the loading of a trap of depth 4 mK. These coefficients were obtained from the loading curves at each repump value with the primary MOT intensity and detuning fixed. We did not see a significant effect on the loading rate, however the two body loss coefficient is clearly seen to decrease by an order of magnitude from $\beta_{L}^{\prime} = 4 \pm 1 \times 10^{-3}$ (atoms s)$^{-1}$ to  $\beta_{L}^{\prime} = 4 \pm1 \times 10^{-4}$ (atoms s)$^{-1}$ as the hyperfine intensity is increased from 5 $\mu$W/cm$^{2}$ to 100 $\mu$W/cm$^{2}$.  The loading rate $R_{0}$ is seen to slightly increase and peak at around 20 $\mu$W/cm$^{2}$. We observe that the loading ate $R_{0}$ is in fact lower at very low hyperfine intensities and that the rate stays higher and constant at higher intensities. The optimal rate was  at hyperfine intensity $I_{R} = 35~\mu$W/cm$^{2}$. And the two body loss coefficient $\beta_{L}^{\prime}$ is seen to decrease with  increasing repump power. All these measurements were made at a MOT intensity, $I_{M} = 28$ mW/cm$^{2}$ and detuned $2.5\gamma$ from resonance. During loading the MOT field gradient was 4 G/cm. Within the limits of primary MOT laser detuning that we were able to achieve, no significant improvement in loading was observed.

\begin{figure}[h]
\includegraphics[width=60mm, height=40mm]{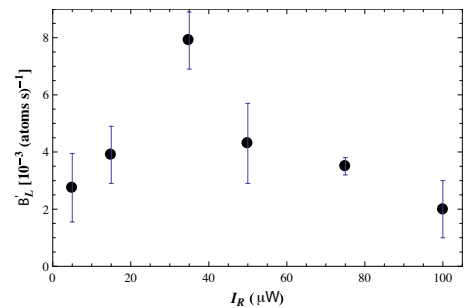}
\caption{The loss coefficient $\beta_{L}^{\prime}$ as a function of repump intensity $I_{R}$ obtained from the number of atoms in the FORT vs holding time  in the presence of the optical molasses. MOT intensity $I_{T} = 28$ mW/cm$^{2}$, FORT depth at 4 mK, P = 2.57 W, $w_{0}= 5.59`\mu$m. Error bars are estimated from the fit.} {\label{fig:betaLTB}}
\end{figure}

The $\beta_{L}^{\prime} $ values that are obtained from the holding curves, with the atoms exposed to the primary MOT and hyperfine light, but without the magnetic field gradient as described in Section \ref{holding}, indicate higher loss rates compared to those obtained from the loading curves. These loss coefficients are shown in Fig.\ \ref{fig:betaLTB}. We expected these values to be similar, within experimental error, to those obtained from the loading curves as has been reported previously in studies of larger dipole traps \cite{20Kuppens:PhysRevA2000}.


\subsection{Loading as a function of Trap depth}
\begin{figure}[h]
\includegraphics[scale=0.5]{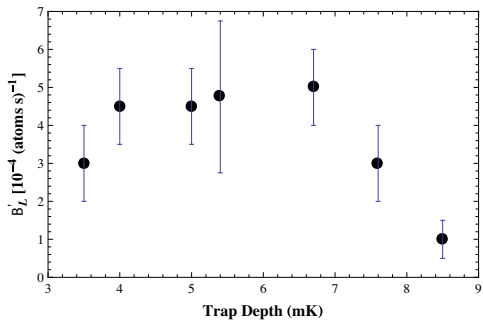}
\caption{Two body loss coefficient $\beta_{L}^{\prime}$ as a function of trap depth. Trap depth is calculated assuming a trap waist of 5.6 $\mu$m. MOT power $I_{M}= 28$ mW/cm$^{2}$, hyperfine repump intensity $I_{R}=35~\mu$W/cm$^{2}$.} {\label{fig:betavsdepth}}
\end{figure}

The FORT was loaded at trap depths ranging from 3.5 mK to 8.5 mK and the loading curves were measured at the optimal hyperfine repump power determined for the 4 mK trap depth of $I_{R} \approx 35 ~\mu$W/cm$^{2}$. The loading rate $R_{0}$ increases from approximately $1.2 \pm 1\times 10^{4}$ atoms/s at 3.5 mK to $2.5 \pm 0.2\times 10^{4}$ atoms/s at 8 mK. The two body loss coefficients $\beta_{L}^{\prime}$ extracted from the fits to the loading curves are shown in Fig.\ \ref{fig:betavsdepth}. 
Overall the two body loss-coefficient decreases as the trap depth is increased.\\

\begin{figure}[h]
\includegraphics[width=60mm, height=40mm]{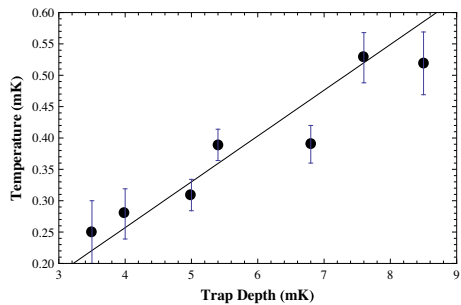}
\caption{Atom temperature in the FORT as function of the trap depth for trap waist $w_{0} \approx 5~\mu$m. A linear fit to the data gives a slope of $T/T_{trap} =0.07\pm 0.04$} {\label{fig:tempvsdepth}}
\end{figure}

The temperature of the trapped atoms was measured by imaging the expanding cloud with the MOT light. The transverse Gaussian  profile $\sigma_{x,y}$ of the cloud increases as $\sqrt{\sigma_{0}^{2}+(k_{B}T/m)t^{2}}$ \cite{20Kuppens:PhysRevA2000}. We did not explicitly verify the temperature in the longitudinal direction. Temperature increases linearly with the trap depth as shown in Fig.\ \ref{fig:tempvsdepth}.
It should be noted that these temperature estimates consist of large uncertainties and indicate only the transverse temperature. We are presently investigating means to make more accurate measurements of the temperature. 

\begin{figure}[h]
\includegraphics[width=60mm, height=40mm]{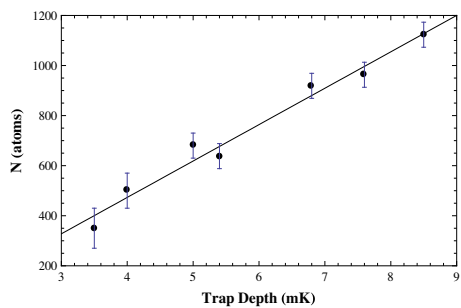}
\caption{Number of atoms in the FORT as a function of trap depth. $w_{0} \approx 5 \mu$m, $I_{MOT}= 28$ mW/cm$^{2}$, $I_{R} = 35~ \mu$W/cm$^{2}$} {\label{fig:atomsvsdepth}}
\end{figure}

\begin{figure}[h]
\includegraphics[width=60mm, height=40mm]{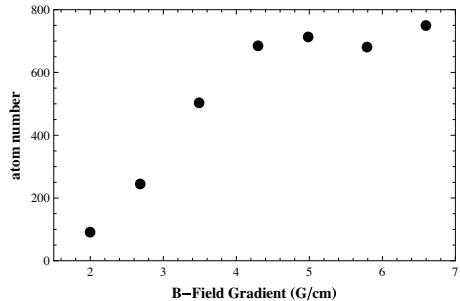}
\caption{Number of atoms in the FORT as the magnetic field gradient is increased $w_{0} \approx 5 \mu$m, trap depth of 4 mK, $I_{MOT}= 28$ mW/cm$^{2}$, $I_{R} = 35~ \mu$W/cm$^{2}$} {\label{fig:atomsvsB}}
\end{figure}

The number of atoms loaded in the trap increase linearly with the trap depth as shown in Fig.\ \ref{fig:atomsvsdepth}. We also note that the loading time to reach the maximum number of atoms decrease from approximately 200 ms to 120 ms as the trap depth increases from 3.5 mK to 8.5 mK. 
We also investigated the loading as a function of the magnetic field and observed that the number loaded gradually increasing and levels off as shown in Fig.\ \ref{fig:atomsvsB}. We assume that with increasing field gradient the MOT is compressed, resulting in an increase of MOT density. And at higher field gradients the MOT density saturates due to reduction in the loading rate of atoms in to the MOT. Therefore the result observed here may not be due to FORT loading properties per-se. 
\section{Conclusion}
We find that it is possible to load a few hundred to a thousand atoms in a microscopic dipole trap.  Trap is loaded most efficiently at higher hyperfine intensities compared to the hyperfine intensities that optimize loading larger and shallower traps. We see no significant improvement in loading within the range of detuning of the main MOT laser that we were able to realize in this investigations. The atom temperature in the deep microscopic trap is found to be $\approx U_{0}/14$, where $U_{0}$ is the trap depth maximum in mK. This is a lower ratio than has been reported for atoms trapped in shallow wells having larger beam waists. Again we note that the temperature is derived from measurements in the transverse dimensions of the trap, longitudinal temperature is needed to make a definitive conclusion.  We also see a large discrepancy between the loss rates evaluated form the loading curves and those obtained from the lifetime curves, by holding the atoms in the FORT in the presence of the optical molasses. The rates obtained from the loading curves (magnetic field gradient present) were lower than those obtained from the lifetime curves (no magnetic field gradient). The values from the two methods were expected to be the same, as has been reported in studies of larger and shallower traps \cite{20Kuppens:PhysRevA2000}. 

As the MOT density is increased by increasing the magnetic field gradient the number of atoms loaded also increase and then gradually levels off. Further investigations are needed to conclude if the effect is directly related to a FORT loading dynamics that is independent of the MOT dynamics. 

In summary, we have shown that it is possible to attain densities as high as $10^{13}$ atoms/cm$^{3}$ in deep microscopic dipole traps. We anticipate that with further cooling of the atoms that it is possible to reach the limits of density that are required for studies seeking to observe strong localization of photons in a dense ultracold  collection of atoms \cite{Kaiser:PhysRevLett.2008, Havey:LaserPhysLett:2006}.  

Authors gratefully acknowledge the financial support from Hobart \& William Smith Colleges and its  Summer Research Internship program. We thank Charles Sukenik for many helpful discussions and for his comments during the preparation of this manuscript.

\newpage
\bibliography{/Users/admin/Documents/Documents/Master}

\end{document}